\documentclass[12pt]{article}
\makeatletter
\renewcommand{\section}{\@startsection{section}{1}{0pt}%
{-3.5ex plus -1ex minus -.2ex}{2.3ex plus.2ex}%
{\normalfont\large\bfseries}}
\def\postsection{.\@postskip@}
\makeatother

\usepackage{amsmath,amssymb}
 \usepackage{epsfig}
  \usepackage{graphicx}

\def\eV{\relax\ifmmode{\rm e\kern-0.12em V}\else{\rm e\kern-0.12em V}\fi}
\def\MeV{\relax\ifmmode{\rm M\eV}\else{\rm M\eV}\fi}
\def\GeV{\relax\ifmmode{\rm G}\eV\else{\rm G\eV}\fi}
\def\as{\relax\ifmmode \alpha_s\else{$\alpha_s$}\fi}
\def\al{\relax\ifmmode\alpha\else{$\alpha$}\fi}
\def\albar{\relax\ifmmode{\bar{\alpha}}\else{$\bar{\alpha}${ }}\fi}
\def\albars{\relax\ifmmode{\bar{\alpha}_s}\else{$\bar{\alpha}_s${ }}\fi}
\def\alps{\relax\ifmmode\alpha_s\else{$\alpha_s${ }}\fi}
\def\msbar{\relax\ifmmode\overline{\rm MS}\else{$\overline{\rm MS}${ }}\fi}
\def\alE{\relax\ifmmode\al_E\else{${\alpha}_E${ }}\fi}
\def\Agoth{\relax\ifmmode{\mathfrak A}\else{$\,{\mathfrak A}${ }}\fi}
 \def\Agothk{\relax\ifmmode{\mathfrak A}_k\else{${\mathfrak A}_k${ }}\fi}
 \def\Acal{\relax\ifmmode{\cal A}\else{${\cal A}${ }}\fi}
\def\Acalk{\relax\ifmmode{\cal A}_k\else{${\cal A}_k${ }}\fi}
 \newcommand{\beq}{\begin{equation}}
\newcommand{\eeq}{\end{equation}}

\begin{document}
\begin{center}

 {\LARGE\textbf{Large regular QCD coupling \\
    at Low Energy ?}}
\vskip4.5mm

{\large Dmitry SHIRKOV}  \vskip2mm

\textit{Bogoliubov \ Lab,\ JINR \ Dubna}  \vskip7mm

\textbf{Abstract} \vskip2mm

\parbox[t]{120mm}{\small  The issue is the expediency of the QCD
 notions use in the low energy region down to the confinement scale,
 and, in particular, the efficacy of the QCD invariant coupling \
 $\albars(Q^2)\,$ with a minimal analytic modification in this domain.
 To this goal, we overview a quite recent progress in application of the
 ghost-free Analytic Perturbative Theory approach (with no adjustable
 parameters) for QCD in the region below 1 GeV. Among them the
 Bethe-Salpeter analysis of the meson spectra and spin-dependent
 (polarization) Bjorken sum rule.\par
  The impression is that there is a chance for the theoretically
 consistent and numerically correlated description of hadronic events
 from $Z_0\,$ till a few hundred MeV scale by combination of analytic
 pQCD and some explicit non-perturbative contribution in the spirit
 of duality.\par
  This is an invitation to the practitioner community for a more
 courageous use of ghost-free QCD coupling models for data analysis
 in the low energy region.} \end{center}\medskip

\section{The pQCD overview}

 {\bf QCD effective coupling \albars.} Common perturbative QCD (pQCD)
 based upon Feynman diagrams starts with power expansion in
 $\as=g_s^2/4\pi\sim 0.1-0.4\,,$ the strong
 interaction parameter analogous to the QED fine structure constant.

   In QFT, an important physical notion is an invariant (or effective,
 or running) coupling function $\albar(Q)\,,$ first mentioned in the
 QED context by Dirac (1933). In the current practice it was
 introduced in the basic renormalization group papers of the
 mid-50s\cite{BSh:55-56}.
 \vspace{1mm}

 The one-loop invariant QCD coupling sums up leading order (LO) logs
 into a geometric progression (with the Bethke\cite{Bethke:06}
 convention for the $\beta_k\,$ coefficients)
 \begin{equation}\label{alQCD-1}
 \albars^{(1)}(Q)= \frac{\as(\mu)}{1+\as(\mu)\beta_0 \,
 \ln(\tfrac{Q^2}{\mu^2})}=\frac{1}{\beta_0\, L},\quad L=
 \ln\left(\tfrac{Q^2}{\Lambda^2}\right)\, ; \quad
 \beta_0=\frac{33-2\,n_f}{12\pi} >0 \,.\eeq

  At high enough energy (small distance), the QCD interaction
 diminishes $ \albars(Q)\sim 1/\ln Q \to 0\quad \mbox{as} \quad
 Q/\Lambda\to\infty;\quad  r\,\Lambda\to 0\,.$ This feature is the
 famous phenomenon of Asymptotic Freedom. 

 At the same time, eq.(1) obeys unphysical singularity (Landau pole)
 $\sim 1/(Q^2-\Lambda^2)$ in the low-energy physical region at
 $|Q|=\Lambda\sim 400\,\MeV\,.$  Transition to the 2-loop case does
 not resolve the issue. \par

  The asymptotic freedom behavior $1/\ln Q $ remains dominant in the
 2-loop or Next-to-Leading-Order (NLO) case. Here, an explicit
 expression for \albars obtained by iterative approximate
 solving\cite{BSh:55-56,Sh-TMP:81} of differential RG equation can
 be written down in a compact the ``denominator form" (as it was
 recently motivated in \cite{den06})
  \begin{equation}  \label{alQCD-2}                 
 \albars^{(2)}(Q)=\frac{1}{\beta_0\,L+\frac{\beta_1}{\beta_0}\ln L}
 \,;\quad\,\beta_1(n_f)=\frac{153-19n_f}{24 \pi^2}\,\end{equation}
 with values
 $ \beta_0(4\pm1)=0.663\mp 0.053\,;\,\beta_1(4\pm1)=0.325\mp 0.085\,.$

 The QCD scale in the $\msbar\,$ scheme $\Lambda^{(n_f)}=
 \Lambda^{(n_f)}_{\msbar}\,,$ as obtained from the data 
 happens to be close to the confinement scale
  $ \Lambda^{(4\pm1)}\sim 300\mp 100\,\,\MeV \simeq 2\,m_\pi$ or
  $R_\Lambda \sim 10^{-13}\,\mbox{cm}.$
\begin{figure}[th]                                     
 \centerline{\includegraphics[width=0.55\textwidth,height=0.40\textwidth]{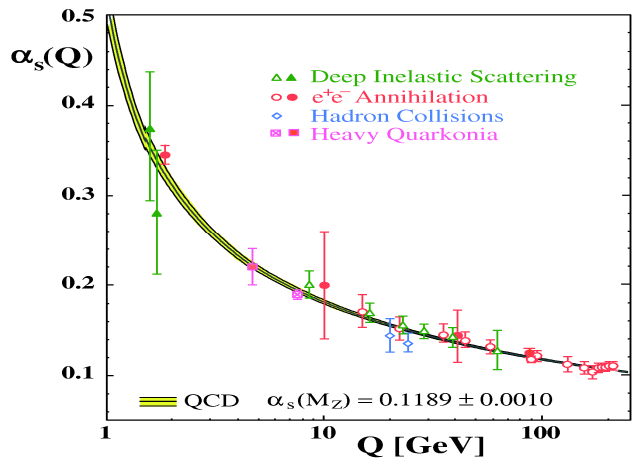}}
 \caption{\sl{\small Effective QCD coupling correlating all the data
 in the range from a few \text{GeV} \ up to a few hundred \text{GeV}.
 The solid curves correspond to the 2-loop, NLO case. Taken from the
 Bethke paper \cite{Bethke:06}.}}
 \end{figure}

  According to Bethke\cite{Bethke:06}, the 2-loop pQCD approximation
 (\ref{alQCD-2}) turns out to be sufficient for numerical correlation
 of several dozen of various experiments. Indeed, Fig.1 gives the
 evidence for {\it the two-loop pQCD triumph}: the NLO theoretical
 curve describes quite accurately - within the current experimental
 and theoretical errors -- all the data in the energy range from 5
 up to a few hundred \GeV.

 However, below 5 \GeV \ the correlation is not so persuasive. Moreover,
 in this region the data on Fig.1 (as well as in the corresponding PDG
 \cite{pdg06} plot) are rather scanty. The reason is not a shortage of
 experiments but rather troubles of their theoretical analysis. Among the
 latter -- the issue of unphysical singularities, like the ``Landau pole".

  As it is well known, the widely-used expressions for effective QCD
 coupling (like eqs.(\ref{alQCD-1}),(\ref{alQCD-2}); see also eq.(7)
 in Ref.\cite{Bethke:06}) and eq.(9.5) in Ref.\cite{pdg06}) suffer
 from spurious singularities in the LE physical region at
 $|Q|\sim\Lambda^{(3)}\sim 400\,\MeV.$  This trouble is one of the main
 embarrassments for the data analysis by pQCD theory below a few \GeV.
 \smallskip

 {\sf Unphysical pQCD singularity vs. lattice data.} At the same time,
 numerous lattice simulation results \cite{as01} -- \cite{skw01}
 testify to the regularity of $\albars(Q)$\ behavior in the region
 below 1 \GeV. Indeed, as it was summarized in papers
 \cite{fourier02,p+r+s:07}, all the lattice data indicate smooth
 growth of \as till specific scale
 $Q= Q_*\sim 400-500\, \MeV\,$ (that is close to $\Lambda^{(3)}\,$)
 with typical values $\albars(Q_*)\sim 0.5-0.8<1\,.$ 

 This means that \underline{\sf common iterative solutions} of RG eqs.,
 like (1), (2) not only can but \underline{\sf should be modified in
 low-energy region} to get rid of singularities and correlate with
 lattice data. \medskip

  {\sf Modifications of ``Common pQCD" in the LE domain.}  
 Several attempts to elude the pQCD singularities have been undertaken
 since the 80s. Among them are the straightforward freezing
 \cite{grun80-82} and a few other, more sophisticated, like glueball
 mechanism \cite{simon89} and exponential modification \cite{cvetic07}.
 All of them introduce some model parameters.\par

  Meanwhile, in the mid-90s, an elegant way (free of additional
 parameters) to resolve this issue was proposed by
 Solovtsov\footnote{Prof. Igor Solovtsov deceased on July 28, 2007.}
 and collaborators \cite{js:95} -- \cite{Sh-Sol:97} on the basis of
 the causality principle implemented in the form of the K\"allen --
 Lehmann analyticity for the QCD coupling $\albars(Q^2)\,.$ Then, on
 the ground of $Q^2$-analyticity, a consistent scheme known as
 Analytic Perturbation Theory (=APT) has been elaborated \cite{MS:97}
 -- \cite{Sh-tmp:01} during the last decade. \par

 Below, we give resume of the APT essence (Sect.2) and its application
 to data (Sect.3) in the above-mentioned troublesome region. These
 results rise hopes that the Bethke's {\it issue of two-loop \as \
 adequacy} can be proliferated to one more order of magnitude -- down
 to a few hundred \MeV \ with the help of analytically modified QCD
 coupling and some additional nonperturbative means in the spirit
 of duality.

 \section{\large Analytic Perturbation Theory}
 Here, we start with a sketch of APT. For details see the review
 papers \cite{Sh-Sol:99} --  \cite{cvetRev08}.

 \subsection{APT - General}                   
  As it is well known, the \underline{1st step} of improving a
 renormalized PT result is supplied by the RG Method \cite{BSh:55-56}
 which allows one to reveal the correct structure of the singularity
 of a partial solution; in the QFT case -- the correct UV and IR
 asymptotics. Its essence is a technique of restoring the so-called
 {\it renormalization\footnote{Or, more exactly, by the {\it
 reparameterization invariance} \cite{Sh:82} of a partial solution.
 Recently, this RG technique has been devised for a class of boundary
 value problems of classical mathematical physics\cite{Sh-Kov:06}.}
 invariance}. In QFT, the RG-improved results
 obey a drawback, the unphysical singularity. \medskip

  In the latter case, the \underline{2nd step}, a further improving of
 RG-invariant PT solution should be used. Its main idea, imposing of
 the {\it analyticity imperative} that in turn stems out of the
 causality condition, was first formulated in the QED context
 \cite{BogLogSh:59}. A more elaborate QCD counterpart, the APT
 algorithm, is based on the following principles :
  \begin{itemize}
 \item \underline{Causality,} that results in the analyticity of the
 effective coupling in the complex $Q^2\,$ plane a l\`a
 K\"allen-Lehmann representation\footnote{For some cases it is
 implemented in a form of the
 Jost-Lehmann (see Sec.4 in Ref.\cite{Sh-Sol:99}) representation.}
{\small  \[ \label{spectral}
 \albars(Q^2)\to \al_{\rm E}(Q^2)= \frac{1}{\pi}\int_0^\infty
 d\sigma\,  \frac{\rho(\sigma)}{\sigma+Q^2- i \epsilon}\,.\]}
  This property provides the absence of spurious singularities.
 \item \underline{Correspondence} with perturbative RG-improved input
 by proper defining \ $\rho(\sigma)=\mbox{Im}\,\albars(-\sigma)\,.$ 

 \item \underline{Representation invariance}, i.e., compatibility
 with linear integral transformations, like a transition from the
 Euclidean, transfer momentum, picture to the Minkowskian, c.o.m.
 energy, one:
 \[\alpha_E(Q^2)=
 Q^2\int^{\infty}_0\frac{\alpha_M(s)\,d s}{(s+Q^2)^2}\,\]
 (or the Fourier transition from $\alpha_E(Q)$ to its Distance image
 $\al_{\rm D}(r)$) that yields \cite{Sh:TMP99} non-power functional
 expansions for observables -- see, below eqs.(\ref{eq4}),(\ref{eq5}).
 \end{itemize}\vspace{-5mm}  

 \subsection{The APT Algorithm }                   
 {\bf Euclidean functions.} Euclidean ghost-free expansion
 functions \cite{Sh:TMP99}  are defined
 \begin{equation}\label{AAPT}
\Acal_{n}(Q^2)=\int_0^\infty\frac{\rho_n(\sigma)}{\sigma
 +Q^2}\,d\sigma, \quad \rho_{n}(\sigma) = \frac{1}{\pi}\,
 \mbox{Im}\!\left[\albars(-\sigma- i\varepsilon)\right]^n \eeq
 via powers of \albars. They form a nonpower set of functions
 $\{\Acal_k(Q^2)\}\,$ that serves as a basis for modified non-power
 APT expansion of RG invariant objects in the $Q\,$ picture, like
 the Adler D-function. The first of these functions can be treated
 as an Euclidean APT coupling
 $\alpha_{\rm E}(Q^2)=\Acal_1(Q^2)\,.$  In the one-loop case
 \[ \alpha_{\rm E}^{(1)}(Q^2) = \frac{1}
 {\beta_{0}}\! \left[\frac{1} {\ln(Q^2/\Lambda^2)}+
 \frac{\Lambda^2} {\Lambda^2-Q^2}\right]\!\]
 it differs from the usual one $\alps(Q^2)$ by the term {
 $\sim 1/(\Lambda^2-Q^2)\,$} that subtracts the singularity. \par
  Here, higher expansion functions are related by the elegant
 recurrent relation
 \[\mathcal{A}^{(1)}_{n+1}(Q^2) = -\frac{1}{n\beta_0}
  \frac{d\,\mathcal{A}^{(1)}_{n}(Q^2)}{d\,\ln Q^2}\,.\]

 {\bf Minkowskian expansion functions} \ are connected
 \cite{js:95,MS:97,Sh:TMP99} with the Euclidean ones
 by contour integral and the reverse ``Adler transformation"
 \[\label{R-oper}
 \Agothk(s)=\frac{i}{2\pi}\,\int^{s+i\varepsilon}_{s-i
 \varepsilon}\frac{dz}{z}\,\mathcal{A}_k(-z)\,;\quad\Acal_k(Q^2)
  = Q^2\int^{\infty}_0\frac{\,\Agothk(s)\,d s}{(s+Q^2)^2}\,.\]

  The Minkowskian APT coupling $\alpha_M(s)=\Agoth_1(s)\,$ in the
 1-loop case
 {\small \[\alpha_M^{(1)}(s)= \left.\frac{1}{\pi\beta_0}
 \arccos\frac{L} {\sqrt{L^2 +\pi^2}}\right|_{L>0}=
 \frac{1}{\pi\,\beta_0}  \arctan\frac{\pi}{L}\,,\quad
 L=\ln (s/\Lambda^2)\]}
 is quantitatively close to the Euclidean APT one; see Fig.2.  
\begin{figure}[ht] \vspace*{-1mm}                    
$${\epsfig{file=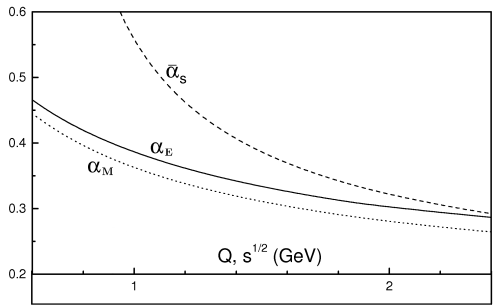,width=0.5\textwidth,angle=0}}$$
\vspace*{-14mm}

\begin{center}
 \parbox{12.6cm}{\caption{\sl{\small Comparison of singular
 $\bar{\alpha}_s$ coupling with Euclidean $\alpha_\text{E}$
 and Minkowskian $\alpha_\text{M}$ in a few \text{GeV} region.
  Taken from paper \cite{ShS_PL:98}.}
 \label{fig2} }}
 \end{center} \end{figure}   

\medskip 

 {\bf Non-power APT - Loop and RS Stability.} In APT, instead of
 universal power-in-\albars ($\albars(Q^2)\,$ or $\albars(s)\,$) series
 \[ d_{pt}(Q^2/s)= d_1\albars(Q^2/s)+ d_2\,\albars^2+ 0(\albars^3)\,,\]
 one should use for each representation its own particular non-power
 expansion
 \beq\label{eq4}
 d_{\rm an}(Q^2)=d_1\,\alpha_E(Q^2) +d_2\,{\cal A}_2(Q^2)+
  d_3\,{\cal A}_3(Q^2)+\dots\,\,,\eeq
  \beq\label{eq5}
 \phantom{aaa}r_{\pi}(s)= \ d_1\,\alpha_M(s) \ + \ d_2\,\Agoth_2(s)
  \ + \  d_3\:\Agoth_3(s) \ +  \dots\,\,\eeq
 that provides better loop convergence and practical RS independence
 of observables.\\
 The 3rd terms in (3), (4) contribute to observables less than 5 \%
 \cite{Sh:01}. Again the 2-loop (NLO) level is practically sufficient.

\smallskip

{\bf Fractional APT}. \ In computation of higher-order corrections
 to inclusive and exclusive processes one deals with non-integer
 {fractional} powers of QCD coupling. In such a case, special {\it
 fractional} generalization has been devised\cite{bks05-12} and
 successively applied to pion form factor \cite{bms05-11} and to the
 Higgs boson decay into a $b\bar{b}$ pair \cite{bms07-7}.

  \subsection{ APT functions at LE region}       
 Comparison of APT Euclidean $\al_E\,$ and Minkowskian
 $\al_M\,$ couplings reveals that below 2-3 \GeV \ scale they,
 being close to each other, differ seriously from the common
 singular \albars -- see Fig.2. Qualitatively, the same is
 true for higher expansion functions.\smallskip

 {\bf The APT RenormScheme- and loop- stability.} \ In Fig.3,
 we give Euclidean APT coupling in the one-, two- and three-loop
 (NNLO) approximations taken in the \msbar scheme.\vspace{5mm}

 \begin{figure}[h] \unitlength=1mm                 
  \begin{picture}(20,30)
  \put(28,-3){\epsfig{file=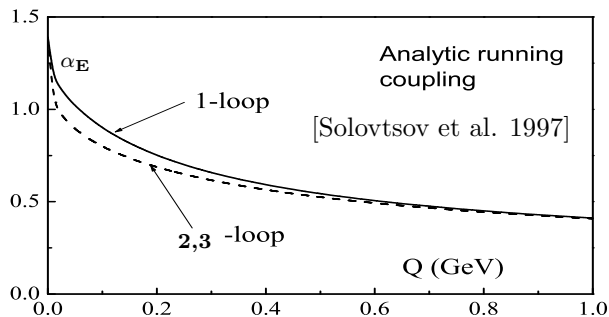,width=7.9cm,height=41mm,angle=00}}
  \put(35,30) {\bf\scriptsize $\alpha_\text{E}\,$}
   \put(50.4,6.2){\scriptsize\bf 2,3}
  \put(68,21) {\footnotesize [Solovtsov et al. 1997]}
  \end{picture} \vspace*{1mm}
  \begin{center}
 \parbox{12.6cm}{\caption{\sl{\small Euclidean APT coupling $\alpha_E$
 in the 1-, 2- and 3-loop cases for the $\overline{\rm MS}$ scheme.
 Taken from paper \cite{ShS_PL:98}.}}}  \end{center}
 \end{figure}
  A beautiful feature of these curves is their relative loop stability.
 The two-loop curve below 1 \GeV \ differs from the three-loop one by
 less than 3 per cent. Hence, the APT two-loop (NLO) curve is accurate
 enough for practical use at three-flavor region. This correlates with
 the Bethke's thesis for higher energies.\par

  In a real QCD case, one has to take into account the proper
 conjunction of regions with different values of effective flavour
 number $n_f\,.$ This, a bit subtle issue was elaborated in
 \cite{Sh-tmp:01}. From the practical point of view, one should use
 common matching conditions for recalculation of $\Lambda^{(n_f)}$
 values at the quark threshold crossing.
 Resulting Euclidean functions $\Acalk$ turn out to be smooth
 in the threshold vicinity, while Minkowskian ones $\Agoth_k$
 remaining continuous have jumps in derivatives.

 Recall here that all this is valid for simple APT functions {\it
 without additional parameters.} This version is known as a {\it
 minimal APT}. Below, we shortly mention its massive generalization
 which contains an additional fitting parameter.  \medskip

 {\bf The ``massive"\ APT modification.} A quite natural ansatz has
 been added to the minimal APT formalism in \cite{nest05}. There, the
 lower limit in the K\'allen-Lehmann integral Eq.(\ref{AAPT}) was
 changed from zero to $m^2>0\,.$  This parameter reminiscent of pion
 mass $m_\pi\,$ squared can be used for the data fitting (as in
  Fig.4).

\section{Low energy APT application }
  The APT approach during the decade of it existence has been applied
 to a number of low energy (above 1-2 \GeV) hadronic observables.
 One has to mention here sum rules \cite{MSS:99,MSS_PL:98},
 $e^+\,e^-\,$ inclusive hadron annihilation \cite{e+e-}, $\tau\,$
 \cite{MilOlS:97} and $\Upsilon\,$ decays\cite{Ups-decay},
 above-mentioned formfactors\cite{bms05-11,bms07-7}, and some others.
 For details one could address to review papers \cite{Sh:01,Sh-Sol:07}.
 Below, we shortly overview quite fresh APT applications to processes
 in a rather low-energy region $\lesssim 1\, \GeV.$

\subsection{APT and bosonic spectrum}      
 {\bf APT + Bethe-Salpeter formalism.}\ Here, is a summary of recent
 analysis \cite{MilanPRL:07} of the meson spectrum by combination of
 the Bethe-Salpeter equation for the $(q,\bar{q})\,$ system with
 the APT approach.

 By use of the 3-dimensional reduction, the BS eq. takes the
 form of an eigenvalue equation for a squared bound state mass
 $$ M^2 = M_0^2 + U_{\rm OGE}+U_{\rm Conf}\,,$$
 with {\small $M_0 =\sqrt{m_1^2+{\bf k}^2}+\sqrt{m_2^2 + {\bf k}^2}$ }
 -- kinematic term,\ $U_{\rm Conf}$ -- {\it confining potential},
 $U_{\rm OGE}$ -- one-gluon exchange potential $\sim$ QCD coupling
 \[\langle {\bf k} \vert U_{\rm OGE} \vert {\bf k}^\prime
 \rangle= {\alps}({\bf Q}^2)\, M_{\rm OGE}({\bf Q=
  k-k^\prime \,, k })\,.\]
 For a given bound state $a\,,$ one has \ (for details see
 Refs.\cite{MilanPR:07})
  \[m^2_a=\langle\phi_a|M_0^2|\phi_a\rangle+ \langle\phi_a|
 U_{\rm OGE}|\phi_a\rangle+ \langle\phi_a|U_{\rm Conf}| \phi_a\rangle\,. \]
 Last two relations allow one to extract $\alps(Q^2_a)\,$ values for
 a low enough momentum transfer region $100\,\MeV\,< \,Q_a < 1\,\GeV\,$.
 \vspace{-2mm}

\begin{figure}[h]\label{fig4} \unitlength=1mm
 \begin{center}
  \begin{picture}(80,50)\put(-9,-6)%
  {\epsfig{file=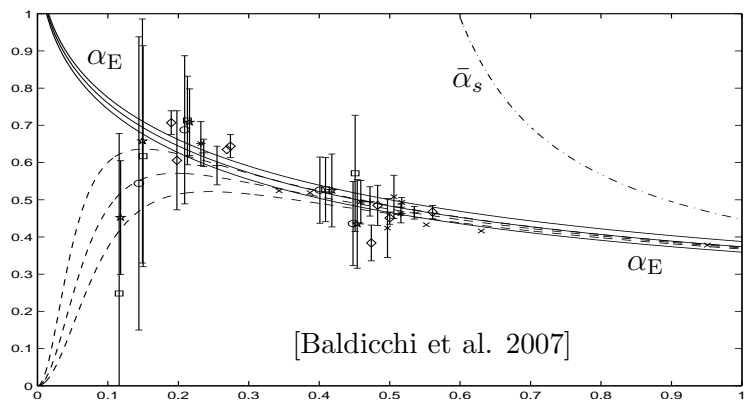,width=0.65\textwidth,height=52mm}}
 \put(1,38){$\alpha_\text{E}$}
  \put(72,11){$\alpha_\text{E}$}
   \put(49,35){$\albars$}
   \put(28,00){\small [Baldicchi et al. 2007]}
   \end{picture}
  \vspace*{5mm}

 \parbox{142mm}{\caption{\sl{\small Comparison of $\alps$ \ from meson
 spectrum (points with error bars) and 3-loop\ {$\alpha_\text{E}$} at
 $\Lambda^{(3)}_{n_f=3}=(417\pm 42)\,$ \text{MeV} (3 solid curves).
 Singular 3-loop $\albars$ coupling  (dot-dashed) is excluded by data.
 Dashed lines correspond to the massive APT version \cite{nest05}
 with $m\sim 40\,\MeV\,.$ Taken from  paper \cite{MilanPRL:07}.}}}
\end{center}\end{figure}

 {\bf Results of \as \ extraction from the bosonic spectrum} are given
 in Fig.4. One sees that meson spectrum data roughly follow a bunch
 of three \alE curves for $\Lambda^{(3)}_{n_f=3}=(417\pm 42)\,$\MeV \
 corresponding to the 2006 world average $\albars(M_Z^2)=
 0.1189\pm0.0010\,.$   There is also a slight hint at the tendency
 for BS-extracted \as values at $Q< 200\, \MeV\,$ to diminish in the
 IR limit. The dashed curves on Fig.4 just relate to this possibility.
 However, in our opinion, such a scenario is supported only by data
 from the D and F
 orbital excitations of the $(q,\bar{q})\,$ system. They have big
 error bars and some of them are subject to a doubtful interpretation.
 If we exclude higher states and limit ourselves to the S and P ones,
 the resulting picture will change.  
\begin{figure}[h]                                  
 \centerline{\includegraphics[width=0.66\textwidth,height=62mm]{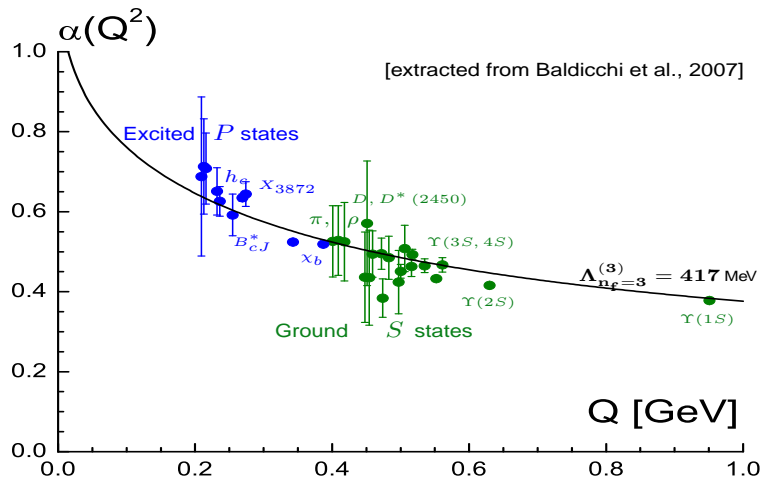}}
 \begin{center}
 \parbox{12.6cm}{\caption{\small The APT $\alpha_\text{E}(Q^2)$
 coupling correlated with the world average vs. $\alpha_s^{exp}$
 from the S,P states of the $(q,\bar{q})\,$ system. Evidence
 for evolution below 500 \text{MeV}. \label{fig5}}}
 \end{center} \end{figure}

 {\sf APT vs S and P data.} In Fig.5, we show the picture without
 higher orbital D and F excitations. This limited set of data with
 small error bars quite nicely follows the minimal APT coupling curve
 with the world average $\Lambda^{(3)}_{n_f=3}=417\,$ \MeV \ value.

 \subsection{Bjorken sum rule}
 Fresh 2006 Jefferson Lab data on the Bjorken Sum Rule for the
 moment of the spin-dependent structure function $\,\Gamma_{p-n,1}\,$
 at $\,0.1<Q^2<3\,\GeV^2$ were analysed recently in the NLO
 approximation\cite{pst:08}. Higher twist (HT) values extracted
 within the APT provide evidence for better convergence of HT series
 as compared to the standard pQCD. As a final result, a reasonable
 quantitative description of the data down to $350\,\MeV$ was
 achieved.\medskip

  Together with the meson spectra evidence Fig.\ref{fig5}, this
 produces an impression that minimal APT allows one to enlarge the
 domain of analytic perturbative QCD (supported by  transparent
 non-perturbative elements) description of hadronic events down to
 a few hundred \MeV.

 \section{\large APT \ in \ QCD: \ Conclusion}

  Meson spectrum data analyzed by the Milano BS-technique with the
 one-gluon exchange potential and confinement ansatz result\cite{Milan:05}
 in a possibility to extract the QCD coupling $\albars(Q)$ \ values in
 the LE domain of momentum transfer $Q< 1\,\GeV.$ In a recent
 research it was shown \cite{MilanPRL:07,MilanPR:07} that the use of
 ghost-free analytic QCD Euclidean coupling $\al_E\,$ in this
 analysis yields rather an intriguing correlation (shown in Figs.4
 and 5) of the ``meson spectrum \as \ values" in the region
 $250\,\MeV\lesssim Q< 1\,\GeV\,$  with the world average
 $\albars(M_Z^2)\,.$\par 

  Along with this, the arena for the APT nonpower expansion results
 for the Bjorken sum rule\cite{pst:08} is also ranging down as far
 as to the $\sim300 - 400\,\MeV$  scale.\bigskip

 Both the results  -- \vspace{-2mm}
 \begin{itemize}
\item exclude common \as \ \ singular behavior and smooth ``freezing"
   below 1 \GeV,
\item support minimal APT extension of pQCD, giving hope for a
 quasi-perturbative consistent quantitative picture from 200 \GeV \
 to 200-300 \MeV. 
\end{itemize} 

  Due to this, there appears a chance for the real possibility of
 consistent theoretical analysis of hadronic processes in the
 low-energy region, the chance that is based on two elements: \\
 -- the procedure of getting rid of spurious singularities, by some
 low-energy modification of pQGD, like the APT one; \\
 -- addition of some appropriate nonperturbative elements in the
 spirit of parton-hadron duality, like confinement ansatz and
 higher twist contribution.

 We appeal to the QCD practicing community for a more regular use of
 ghost-free QCD coupling models for data analysis in the low-energy
 region below 1 --2 \GeV. Just in this region theoretical
 errors quite often exceed the experimental ones.

 \bigskip \centerline{\bf Acknowledgements}\smallskip
 The author is grateful to Professor Wolfhart Zimmermann and Prof. E.
 Seiler for hospitality in MPI, Muenchen. The useful discussion with
 Drs. A.Bakulev, S. Bethke, S.Mikhailov, O.Solovtsova, N.Stefanis and
 O.Teryaev is sincerely acknowledged. This work was supported in part
 by RFBR grant 08-01-00686, the BRFBR (contract F08D-001) and RF
 Scientific School grant 1027.2008.2.  \vspace{-2mm}


 \end{document}